\documentclass{article}
\usepackage{caption}
\usepackage{graphicx}
\title{A Guided Tour of Planetary Interiors}
\author{Alexander R. Klotz\\Department of Physics, McGill University \\ 3600 rue University, Montreal, Quebec H3A 2T8, Canada\\klotza@physics.mcgill.ca}

\begin{document}
\maketitle

\abstract{We explore the gravitational dynamics of falling through planetary interiors. Two trajectory classes are considered: a straight cord between two surface points, and the brachistochrone path that minimizes the falling time between two points. The times taken to fall along these paths, and the shapes of the brachistochrone paths, are examined for the Moon, Mars, Earth, Saturn, and the Sun, based on models of their interiors. A toy model of the internal structure, a power-law gravitational field, characterizes the dynamics with one parameter, the exponent of the power-law, with values from -2 for a point-mass to +1 for a uniform sphere. Smaller celestial bodies behave like a uniform sphere, while larger bodies begin to approximate point-masses, consistent with an effective exponent describing their interior gravity.}

\clearpage
This letter considers the dynamics of objects falling along paths constrained to the interiors of planetary bodies, where the mass distributions arising from hydrostatic equilibrium lead to non-trivial gravitational fields.   Two classes of paths are considered: a Euclidean cord connecting two points through the interior, and the brachistochrone path that minimizes the time taken to fall between two points. The internal dynamics are calculated for the Moon, Mars, Earth, Saturn, and the Sun, and compared to a model in which the internal gravitational field scales as a power-law. Paths are assumed to be frictionless and planetary rotation and asphericity is ignored.

The best-known problem of this type is the gravity tunnel through the Earth \cite{cooper}. Under the assumption of uniform density, the gravitational potential becomes harmonic and the dynamics are those of a simple harmonic oscillator. A surprising result of this system is that a cord path takes the same amount of time to fall through, 42 minutes, independent of the length of the cord. The brachistochrone curve through a uniformly dense sphere is that of a hypocycloid, the shape found by rolling a smaller circle inside a larger one \cite{venezian}. Recently, the gravitational dynamics of a falling object inside the Earth were calculated using a realistic internal profile \cite{klotztunnel}, and found to be similar to the dynamics under a constant radial field: it is more accurate to assume that the Earth's gravity is the same strength throughout its interior than to assume that the density is constant. This was explained by the fact that the internal mass profile of the Earth increases roughly with the 1.97 power of radius, nearly cancelling Newtonian gravity. Inspired by that work, this letter extends that analysis to other solar system bodies and attempts to classify the dynamics through a single parameter: the scaling exponent of the internal gravitational field.

We consider the dynamics of a test particle falling through a spherically symmetric gravitational field, where the gravitational field scales with radius $r$ as a power-law with an exponent $\alpha$ to some surface field $g_{o}$ at surface radius $R$. The mass $m$ enclosed within radius $r$ increases towards $M$ at $R$ with the exponent $\alpha+2$, related to the gravitational field through Newton's law,

\begin{equation}
g(r)=g_{o}\left(\frac{r}{R}\right)^{\alpha} \ \ \ \ \ \ \ \ \ \ m(r)=M\left(\frac{r}{R}\right)^{\alpha+2} \ .
\label{eq:alpha}
\end{equation}

There is a certain range of $\alpha$ that is physically meaningful. The value $\alpha=1$ corresponds to a uniformly dense sphere, whereas for $\alpha>1$, density increases with radius and hydrostatic equilibrium is violated. Large values of $\alpha$ may describe a hollow shell of mass or a Dyson sphere. The value $\alpha=-2$ corresponds to Newtonian gravity outside a point-mass, thus $\alpha<-2$ is unphysical (such fields can be found in region of negative-mass material, or in a co-orbiting reference frame around a strong gravitational quadrupole or electric dipole). The density is singular at the origin for all $\alpha<1$, but avoiding this singularity leads to tension with the constraint of hydrostatic equilibrium. The physical values of $\alpha$ are therefore $-2\leq\alpha\leq 1$. The special case of $\alpha=-1$ corresponds to a logarithmic potential and must be treated uniquely, but in practice $\alpha=-1+\epsilon$ is a well-behaved approximation.

Falling from rest at the surface, the velocity $v$ of a test mass $\mu$ at any radius can be related to that radius by conservation of energy. The gravitational potential is found by integrating the field from the zero-point $r_{o}$ to that radius. The zero-point is at the origin for $\alpha>-1$ and at infinity otherwise, although ultimately only the difference in potential is relevant,

\begin{equation}
E_{tot}=\frac{1}{2}\mu v^{2}+\int_{r_{o}}^{r}\mu g(r)dr=\int_{r_{o}}^{R}\mu g(r)dr\ ,
\end{equation}
\begin{equation}
v^{2}=2\int_{r}^{R} g(r)dr=2\int_{r}^{R}g_{o}\left(\frac{r}{R}\right)^{\alpha}dr = \frac{2g_{o}}{\alpha+1}R^{-\alpha}\left(R^{\alpha+1}-r^{\alpha+1}\right)\ .
\label{eq:velocity}
\end{equation}

The time taken to fall along the diameter from $r=R$ to $r=0$ can be found by integrating the reciprocal velocity by the line element, and the time taken to fall to the opposite side is twice this:

\begin{equation}
T_{\textrm{diameter}}=2\int_{0}^{R}\frac{dr}{v}=\sqrt{\frac{\pi}{2\left(\alpha+1\right)}}\frac{\Gamma \left(\frac{1}{\alpha+1}\right)}{\Gamma \left(\frac{\alpha+3}{2\left(\alpha+1\right)}\right)}\sqrt{\frac{R}{g_{o}}} \ .
\label{eq:falltime}
\end{equation}

This closed-form expression is only defined for $\alpha\geq-1$ and the value is known from orbital mechanics for $\alpha=-2$ \cite{tee}, and numerical integration can be used to find the time for the entire range. It is useful to consider this time relative to the uniform density ($\alpha=1$) value, which is the same as the time taken to make half an orbit along the surface: $T_{r} \equiv T_{\rm diameter} /\pi\sqrt{\frac{R}{g_{o}}}$. Over the physical range of $\alpha$, the relative time $T_{r}$ is almost perfectly described by the linear function $T_{r}=0.9+0.1\alpha$.

We now consider a straight cord connecting two surface points, subtending an angle $2\Theta$ (Figure 1). Making an angle $\theta$ with respect to the midpoint between the two points, the radial coordinate of any position along the cord is $r(\theta) = R\cos(\Theta)\sec(\theta)$ and the line element along this path is $ds=R\cos(\Theta)\sec^{2}(\theta)d\theta$. The time taken to fall from $\theta=\Theta$ to $\theta=-\Theta$ along a cord can again be found by integrating the reciprocal velocity by the line element, 

\begin{equation}
T_{\textrm{cord}}=\int_{-\Theta}^{\Theta}\frac{ds}{v(\theta)}=\sqrt{\frac{R}{g_{o}}}\int_{-\Theta}^{\Theta}d\theta\frac{\cos\Theta\sec^{2}\theta}{\sqrt{\frac{2}{\alpha+1}\left(1-\left(\cos\Theta\sec\theta\right)^{\alpha+1}\right)}} \ .
\label{eq:cord}
\end{equation}

This has a known solution for $\alpha=1$ where the cord time is independent of $\Theta$. A solution for $\alpha=0$ has been derived \cite{klotztunnel}. No general solution for arbitrary $\alpha$ was found, thus numerical analysis is required. The cord time can either be found by numerical integration of Equation 5 or the kinematic equation of motion; the latter yields smoother results. 

While the cord path minimizes the distance travelled by the falling object, the brachistochrone path minimizes the time taken to fall between two points. To find the brachistochrone, the time taken to fall along an arbitrary path from point $A$ to point $B$ is written in radial coordinates, following the derivation of the generalized brachistochrone \cite{klotztunnel}. The velocity, known from Equation \ref{eq:velocity} is left un-expanded for compactness,

\begin{equation}
T_{\textrm{brac}}=\int_{A}^{B}{dt}=\int_{A}^{B}{\frac{ds}{v(r)}}=\int_{A}^{B}{\frac{\sqrt{dr^{2}+r^{2}d\theta^{2}}}{v(r)}}=\int_{-\Theta}^{\Theta}{d\theta\frac{r^{2}+r'^{2}}{v(r)}} \ .
\label{eq:brach1}
\end{equation}

Because the integrand $dt$ depends only on $r$ and $\frac{dr}{d\theta}$, it can be minimized using the Beltrami identity. An additional constraint is added, such that the path is horizontal at its deepest point: $\frac{dr}{d\theta}=0$ when $r=R_{d}$,

\begin{equation}
dt-r'\frac{d(dt)}{dr'}=C=\frac{R_{d}}{v(R_{d})} \ .
\label{eq:miner}
\end{equation}
$C$ is a constant. This yields the minimization condition for the brachistochrone curve, at which point the power-law velocity can be substituted,

\begin{equation}
\frac{dr}{d\theta}=\frac{r}{R_{d}}\sqrt{\frac{r^{2}v^{2}(R_{d})-R_{d}^{2}v^{2}(r)}{v^{2}(r)}}= \frac{r}{R_{d}}\sqrt{\frac{r^{2}\left( R^{\alpha+1}-R_{d}^{\alpha+1}\right)-R_{d}^{2}\left( R^{\alpha+1}-r^{\alpha+1}\right)}{\left( R^{\alpha+1}-r^{\alpha+1}\right)}}\ .
\label{eq:brach}
\end{equation}

This differential equation can solved for a relationship between $r$ and $\theta$, and can be integrated to find $T_{brac}$. Again, a general solution is lacking but specific cases have been discussed for $\alpha=1$ \cite{venezian}, $\alpha=0$ \cite{klotztunnel}, $\alpha=-1$ \cite{teelog} , and $\alpha=-2$ \cite{tee}. A numerical method \cite{klotztunnel} is used to calculate the brachistochrone path and time for arbitrary $\alpha$. 

\begin{figure}[ht]
	\centering
		\includegraphics[width=0.4\textwidth]{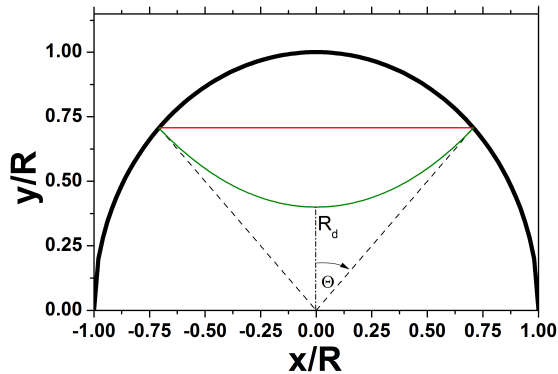}
	\caption{\footnotesize Schematic of the coordinate system used in these derivations. A cord path (red) and brachistochrone (green) subtend an angle $2\Theta$ and the brachistochrone reaches a minimum deepest radius at $R_{d}$.}
	\label{fig:diagram}
\end{figure}

Having derived expressions for falling through a power-law gravitational field along a cord and the brachistochrone, we can examine their behavior as a function of $\alpha$ in Figure 2. First considering the time taken to fall through the diameter, over the physical range of $\alpha$, the relative time ranges from $T_{r}=1/\sqrt{2}$ at $\alpha=-2$ to $T_{r}=1$ at $\alpha=1$, with a near-linear dependence, and continues to increase in the non-hydrostatic regime. The time taken to fall along a cord path is only independent of the surface angle when $\alpha=1$. In other cases it interpolates between this orbital half-period and $T_{\textrm{diameter}}$. In the non-hydrostatic regime, it takes more time to fall along longer paths, as gravitational fields weaken near the origin. Brachistochrone curves, which balance the greatest speeds with the shortest paths, generally become more vertical as $\alpha$ decreases and there is a greater benefit to a longer, deeper path. For larger $\alpha$, the brachistochrone curves begin to approximate a cord, as the shortest path becomes favored.

\begin{figure}[ht]
	\centering
		\includegraphics[width=0.8\textwidth]{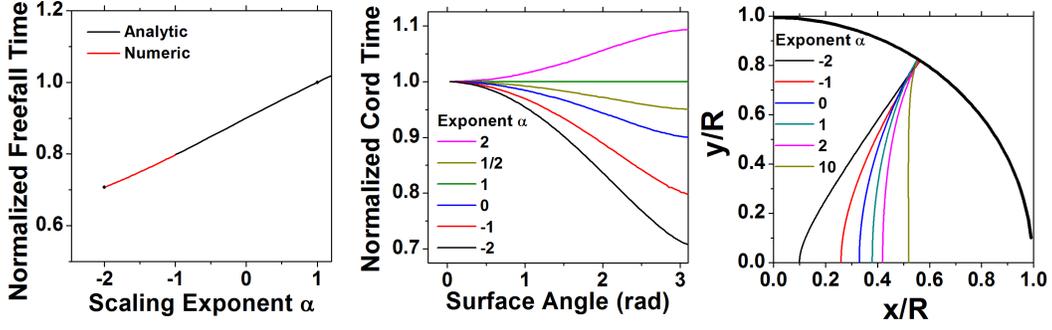}
	\caption{\footnotesize Left: The time taken to fall to the center of a planetary body as a function of the gravity scaling exponent $\alpha$, normalized by the orbital half-period. The black curve is Equation \ref{eq:falltime} and the red curve is a kinematic numerical solution. Middle. Cord fall times as a function of surface angular separation, normalized to the orbital half-period, for several values of $\alpha$. Right. Brachistochrone paths from an arbitrary surface point, for various values of $\alpha$.}
	\label{fig:theory}
\end{figure}
Having considered an ideal case of a power-law gravitational field, we now consider the gravitational dynamics inside Solar System bodies. The interiors of the Moon and Mars are given by a piecewise stratified model taken from Bills and Rubencam \cite{bills}. The interior of the Earth is taken from the Preliminary Reference Earth Model, based on seismic data \cite{prem}. The interior of Saturn is based on a sixth-order polynomial fit of its internal density profile to its gravitational multipole moments, taken from Schubert and Anderson \cite{saturn}, and the interior of the sun is based on polytropic hydrostatic equilibrium from Christensen-Dalsgaard \cite{sun}. The radial mass and gravity profiles of these bodies are seen in Figure 3.

The relative radial mass profiles of the planetary bodies increase from zero at the origin to unity at the surface, although the interpolation is mass-dependent: smaller objects tend to increase more slowly, comparable to a uniformly dense cubic for the moon, while for the larger objects the mass profile is akin to logistic growth as the outer regions contribute little mass.  Planetary bodies that are in hydrostatic equilibrium typically have a gravitational field that is highest in the planetary interior. The Earth, for example, has its strongest gravitational field of 10.8 m/s$^{2}$ at roughly 3000 km below the surface \cite{prem}. Lighter objects such as the Moon and Mars have a gradual near-monotone increase in gravitational field with radius, consistent with small hydrostatic deviations from uniform density. Larger objects, such as the Sun and Saturn, have their mass strongly concentrated towards the center, and have a strong sub-surface gravitational peak and a rapid falloff towards zero at the origin.

\begin{figure}[ht]
	\centering
		\includegraphics[width=0.7\textwidth]{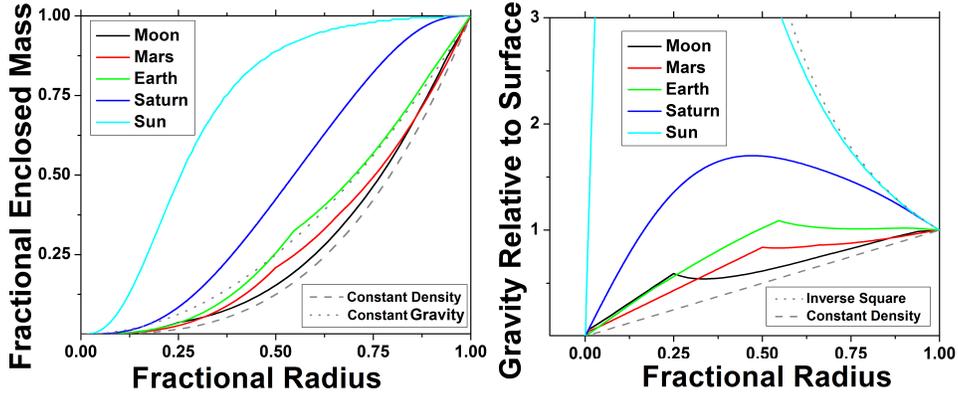}
	\caption{ \footnotesize Left: The radial mass profiles of the Moon, Mars, Earth, Saturn, and the Sun. For comparison, the cases of uniform density (cubic radial dependence) and constant gravity (quadratic radial dependence) are shown. Right: The internal gravitational fields of the Moon, Mars, Earth, Saturn, and the Sun. Uniform density (linear radial dependence) and inverse-square curves are shown for comparison. The curve for the Sun is truncated so that the information in the other curves is visible; it reaches a relative field of 8.6 at a fractional radius of 0.17.}
	\label{fig:Internal}
\end{figure}

\begin{figure}[ht]
	\centering
		\includegraphics[width=0.8\textwidth]{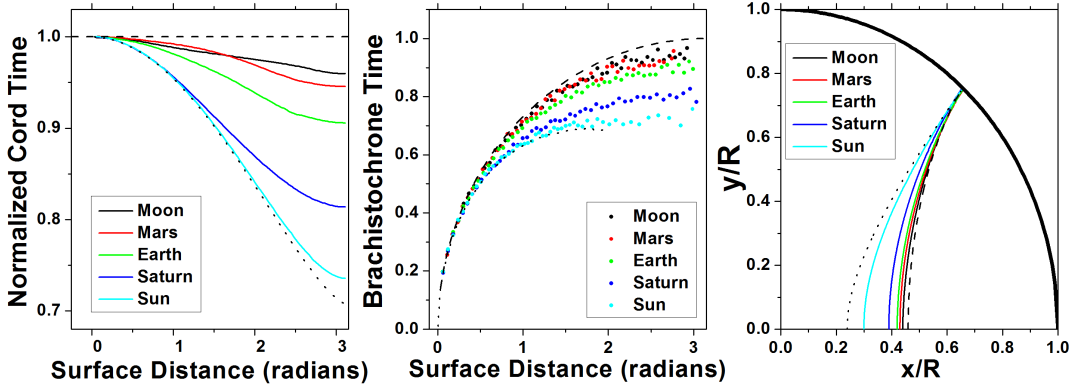}
	\caption{ \footnotesize Left. The cord fall times through the Moon, Mars, Earth, Saturn, and the Sun, as a function of surface angular separation, normalized to the orbital half-period. Middle: The brachistochrone fall times through the Moon, Mars, Earth, Saturn, and the Sun, as a function of surface angular separation, normalized to the orbital half-period. Right: Brachistochrone curves through the Moon, Mars, Earth, Saturn, and the Sun, with the same surface angular separation throughout. In all figures, the behavior for a uniformly dense object (dashed line) and a point-mass (dotted line) are shown for comparison. Note that a brachistochrone curve around a point-mass does not exist for a surface angle greater than $2\pi/3$ \cite{tee}.} 
	\label{fig:Planets}
\end{figure}

The physical properties and absolute diameter fall times can be found in Table 1. The time taken to fall through a cord between two points is shown in Figure \ref{fig:Planets}, left. All curves are bound by the upper limit of uniform density and the lower bound of the point-mass.  Curiously, in addition to being the same angular size as seen from Earth, the Sun and the Moon would take the same time to fall through. Examining the curves in Figure \ref{fig:Planets}, a trend emerges. For smaller objects under hydrostatic equilibrium, gravity deforms the density only slightly, such that the behavior is close to that predicted for uniform density, only decreasing by four percent with angular separation, in the case of the moon. As the object becomes more massive, the behavior approaches that of a point source as more mass is found towards the center. The same trend is evident in the brachistochrone times (Figure 4, center): for smaller objects, the behavior is similar to the prediction of the uniform-density sphere, while for larger objects it approaches that of the point-mass. The brachistochrone paths (Figure 4, right) are shown for an arbitrary surface angle (corresponding to roughly 10,000 km on the Earth), and again this trend is seen, with the curves remaining closer to vertical in larger objects.

\begin{table}[ht]
  \centering
  \caption{ \footnotesize The physical parameters of the celestial bodies, the time taken to fall through their diameters, and the effective power-law exponent $\alpha$ estimated from fits to their mass profiles and from their fall time.}
    \begin{tabular}{|c|c|c|c|c|c|c|}
    \hline
          & Radius (km) & Mass (kg) & $g_{o}$ (m/s$^{2}$) & Diameter Time (min) & Mass $\alpha$ & Time $\alpha$ \\
    \hline
    Moon  & 1,734  & $7.3 \cdot 10^{22}$ & 1.6   & 52    & 0.6   & 0.6 \\
    Mars  & 3,390  & $6.4 \cdot 10^{23}$ & 3.7   & 47    & 0.4   & 0.4 \\
    Earth & 6,380  & $6.0\cdot 10^{24}$ & 9.8   & 38    & 0     & 0 \\
    Saturn & 60,200 & $5.7 \cdot 10^{26}$ & 10.4  & 61    & -0.8  & -0.9 \\
    Sun   & 695,000 & $2.0 \cdot 10^{30}$ & 274   & 52    & -1.5  & -1.7 \\
    \hline
    \end{tabular}%
  \label{tab:addlabel}%
\end{table}%

\begin{figure}[t]
	\centering
		\includegraphics[width=0.8\textwidth]{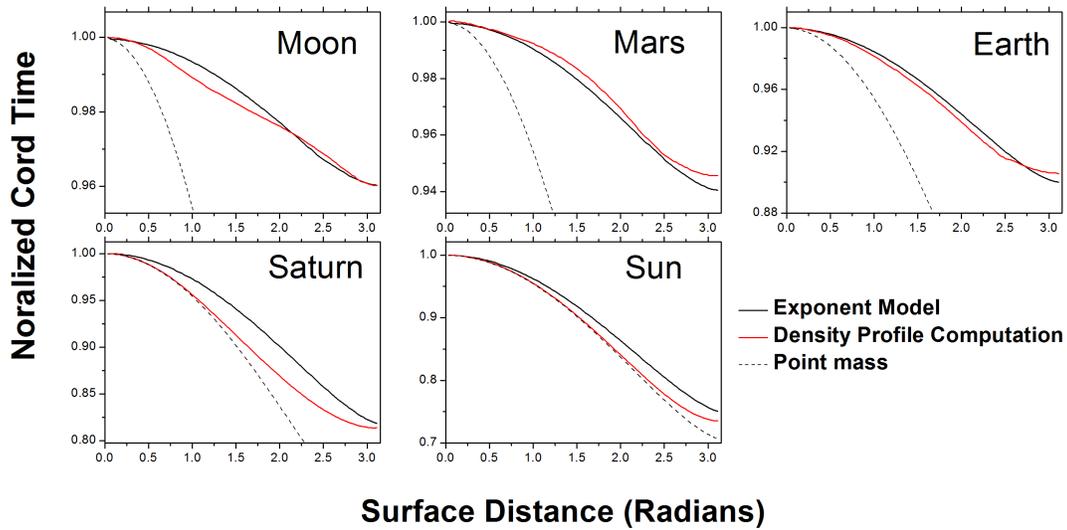}
	\caption{ \footnotesize The line fall times as a function of surface angle for the five discussed bodies, comparing the prediction of a fit exponent to that of the density profile. The case for the point mass is shown for comparison. The y-axis in each figure is different.} 
	\label{fig:fiveplanets}
\end{figure}

We can examine in more detail the line fall times predicted based on the exponents taken from fits to the radial mass profiles (Figure \ref{fig:fiveplanets}). The exponents are found from single-parameter least-squares fits to the function $m(r)=M\left(\frac{r}{R}\right)^{\alpha+2}$. Fits tend to be biased towards the near-surface trends, but most of time spent falling is also near the surface, where velocities are slower. The cord time profiles are qualitatively similar between the exponent model and the prediction from the density profile, with deviations for intermediate surface angles. The Sun is better described by the dynamics around a point-mass than those found from a fit to its mass profile.  To further examine the validity of the power-law model, we can compare an ``effective exponent'' based on the diameter fall time to those calculated from power-law fits to the internal mass profiles, seen in Table 1. The exponents from the two measures are similar in all cases. Overall this shows that despite the non-monotonicity in the internal gravitational fields of these bodies, the power-law model is a good first approximation to their internal dynamics.

In summary, the time taken to fall through straight and minimal paths through the Moon, Mars, Earth, Saturn, and the Sun have been calculated based on their known internal structure and compared to the predictions of a simple power-law model, with good qualitative agreement. A more complex general model can be developed to better characterize the non-monotonicity of internal gravitational fields. The dynamics of rocky material or dust through protoplanetary disks may be characterized by the model developed in this letter, and a natural progression of this analysis is to compute a model for the internal dynamics through a galaxy, comparing the predictions of a dark matter halo to that of a modified gravitational model. The dynamics through a rotating uniformly dense body have been extensively covered by Simoson \cite{simoson}, and an exploration of dynamics through rotating non-uniform bodies may be of interest. 

\textbf{The author acknowledges Grace Dupuis, Chen Karako Argaman, and Kiyoshi Masui for helpful comments.}

\bibliographystyle{unsrt}
\bibliography{guiderefs}

\begin{thebibliography}{10}

\bibitem{cooper}
Paul~W Cooper.
\newblock Through the earth in forty minutes.
\newblock {\em American Journal of Physics}, 34:68, 1966.

\bibitem{venezian}
Giulio Venezian.
\newblock Terrestrial brachistochrone.
\newblock {\em American Journal of Physics}, 34:701--701, 1966.

\bibitem{klotztunnel}
Alexander~R. Klotz.
\newblock The gravity tunnel in a non-uniform earth.
\newblock {\em American Journal of Physics}, 83(3), 2015.

\bibitem{tee}
Garry~J Tee.
\newblock Isochrones and brachistochrones.
\newblock Technical report, Department of Mathematics, The University of
  Auckland, New Zealand, 1998.

\bibitem{teelog}
Garry~J Tee.
\newblock Brachistochrones for attractive logarithmic potential.
\newblock {\em NZJ Math}, 30:183--196, 2001.

\bibitem{bills}
Bruce~G Bills and David~P Rubincam.
\newblock Constraints on density models from radial moments: Applications to
  earth, moon, and mars.
\newblock {\em Journal of Geophysical Research: Planets (1991--2012)},
  100(E12):26305--26315, 1995.

\bibitem{prem}
Adam~M Dziewonski and Don~L Anderson.
\newblock Preliminary reference earth model.
\newblock {\em Physics of the earth and planetary interiors}, 25(4):297--356,
  1981.

\bibitem{saturn}
John~D Anderson and Gerald Schubert.
\newblock Saturn's gravitational field, internal rotation, and interior
  structure.
\newblock {\em Science}, 317(5843):1384--1387, 2007.

\bibitem{sun}
J~Christensen-Dalsgaard, W~D{\"a}ppen, SV~Ajukov, ER~Anderson, HM~Antia,
  S~Basu, VA~Baturin, G~Berthomieu, B~Chaboyer, SM~Chitre, et~al.
\newblock The current state of solar modeling.
\newblock {\em Science}, 272(5266):1286--1292, 1996.

\bibitem{simoson}
Andrew~J Simoson.
\newblock Sliding along a chord through a rotating earth.
\newblock {\em The American Mathematical Monthly}, pages 922--928, 2006.

\end{thebibliography}

\end{document}